\documentclass{PoS}
\usepackage{bm}
\usepackage{epsfig,multicol}
\def\Vec#1{\mbox{\boldmath $#1$}}
\def\dis{\displaystyle}

\def\NPA{{Nucl. Phys.} {\bf A}}
\def\NPB{{Nucl. Phys.} {\bf B}}
\def\PLB{{Phys. Lett.} B}

\def\PRD{{Phys. Rev.} D}

\title{Roles of the quark field in the infrared lattice Coulomb gauge and Landau gauge QCD}

\ShortTitle{Roles of the quark field in the infrared lattice Coulomb gauge and Landau gauge QCD}

\author{\speaker{Sadataka Furui}%
         \thanks{}\\
        Sch of Sci \& Engr; Teikyo Univ, Utsunomiya, 320-8551 Japan\\
        E-mail: \email{furui@umb.teikyo-u.ac.jp}}

\author{Hideo Nakajima\\
        Fac of Infor Sci; Utsunomiya Univ, Utsunomiya, 320-8585 Japan\\
        E-mail: \email{nakajima@is.utsunomiya-u.ac.jp}}

\abstract{The effective coupling of QCD is measured by using the gauge 
configurations produced by the MILC collaboration in which the Kogut Susskind (KS) 
fermion is incorporated and by using that produced by the QCDOC collaboration
 in which the domain wall fermion (DWF) is incorporated. We fix the gauge to the
Landau gauge and to the Coulomb gauge. The infrared effective coupling in the
Coulomb gauge agrees with the recent extraction at JLab, but that in the Landau gauge
 shows infrared suppression. The suppression is expected to be due to
the color anti-symmetric ghost propagator which in the unquenched
configurations has stronger infrared singularity than the color diagonal
ghost propagator.  The Coulomb form factor in the infrared depends on the
kind of the fermion incorporated in the system and the temperature.

The quark has the effect of quenching randomness and the fluctuation of the
color anti-symmetric ghost propagator is reduced in the unquenched configuration, 
and the Kugo-Ojima parameter $c$ is closer to 1 in the unquenched configuration 
than in the quenched configuration.
}

\FullConference{The XXV International Symposium on Lattice Field Theory\\
July 30 - August 4 2007\\
Regensburg, Germany}

\begin{document}
\section{Introduction}
We fixed gauge configurations of the MILC collaboration\cite{MILC96,MILC01} and the QCDOC collaboration\cite{QCDOC07} to  Landau gauge and to Coulomb gauge and studied the 
infrared features of the lattice QCD by comparing the gluon propagator, the ghost propagator, the effective coupling etc. In the case of finite temparature MILC$_{ft}$, we assign the $\beta=5.65, 5.725$ and 5.85 data correspond to $T=143, 172.5$ and 185 MeV$_\rho$, and in the unit of critical temperature $T_c$ by $T/T_c=1.02, 1.23$ and 1.32, respectively.
\begin{table}[htb]
\begin{tabular}{c|c|c|c|c|c|c|c}
   &$\beta$ &N$_f$ &$m$  &$1/a$(GeV)& $L_s$ & $L_t$ &$a L_s$(fm) \\
\hline
MILC$_{ft1}$ &5.65&  2&  0.008   & 1.716 & 24 &12& 2.76 \\
MILC$_{ft3}$ &5.725&  2& 0.008   & 1.914 & 24 &12& 2.47 \\
MILC$_{ft5}$ &5.85&  2& 0.008   & 2.244 & 24 &12& 2.11 \\
\hline
MILC$_c$ &6.83($\beta_{imp}$)&   2+1   & 0.040/0.050& 1.64 & 20 &64&2.41\\
       &6.76($\beta_{imp}$)&  2+1    & 0.007/0.050& 1.64 & 20 &64&2.41\\
\hline
MILC$_f$ &7.11($\beta_{imp}$)& 2+1     & 0.0124/0.031 & 2.19 &28 & 96&2.52\\
       &7.09($\beta_{imp}$)& 2+1     & 0.0062/0.031 & 2.19 &28 & 96&2.52\\
\hline 
MILC$_{2f}$ &7.20($\beta_{imp}$) &2& 0.020& 1.64 & 20 &64&2.41 \\
\hline
MILC$_{3f}$ &7.18($\beta_{imp}$) &3& 0.031 & 2.19 &28 & 96&2.52 \\
\hline
DWF$_{01}$ &2.13($\beta_{I}$)&2+1 & 0.01/0.04 & 1.743(20) &16 & 32 & 1.81   \\
DWF$_{02}$ &2.13($\beta_{I}$)&2+1 & 0.02/0.04 & 1.703(16) &16 & 32 & 1.85  \\
DWF$_{03}$ &2.13($\beta_{I}$)&2+1 & 0.03/0.04 & 1.662(20) &16 & 32 & 1.90  \\
\hline
\end{tabular}
\end{table}

\section{The Lattice Landau gauge QCD}
We measured Kugo-Ojima confinement parameter\cite{KO79} of various configurations in Landau gauge\cite{NaFu00,FuNa06a,FuNa06b}.
The Kugo-Ojima parameter is defined by the two point function of the
covariant derivative of the ghost and the commutator of the antighost and gauge field
\begin{eqnarray}
&&\left(\delta_{\mu\nu}-\frac{q_\mu q_\nu}{q^2}\right)\delta^{ab}u(q^2)\nonumber\\
&&=\frac{1}{V}
\sum_{x,y} e^{-iq(x-y)}\left\langle {\rm tr}\left({\Lambda^a}^{\dag}
D_\mu \displaystyle{\frac{1}{-\partial D}}[A_\nu,\Lambda^b] \right)_{xy}\right\rangle,\nonumber\\ \label{kugou}
\end{eqnarray}
 as $c=-u(0)$. It is a scalar function at vanishing momentum in the continuum theory, but in the lattice simulation, we measured the magnitude of  the right hand side of the eq.(\ref{kugou}) with $\mu=\nu$ polarizations of $A_\nu$ and $D_\mu$, and then observed in the case of asymmetric lattices there is a strong positive correlation between the magnitude and the lattice size of the axis whose directions are perpendicular to the polarization.
In Figure 1 the temperature and polarization dependence of the Kugo-Ojima parameter of MILC$_{ft}$ is presented.

Data in Table 1 show that the parameter $c$ of MILC$_{2f}$ and MILC$_{3f}$ are consistent with 1 while the data of quenched 56$^4$ configuration are about 0.8. 
Using the covariant derivative 
\[
D_\mu(U_{x,\mu})=S(U_{x,\mu})\partial_\mu +[A_{x,\mu},\cdot]
\]
and the parameter $e$ defined as
\[
e=\left\langle\sum_{x,\mu}{\rm tr}(\Lambda^{a\dag} 
S(U_{x,\mu})\Lambda^a)\right\rangle/\{(N_c^2-1)V\},
\]
$h=c-\dis{\frac{e}{d}}$ and dimension $d=4$, it follows that 
$h=0$ in the continuum limit means the horizon condition\cite{Zw91}.
\begin{table}[htb]
\begin{tabular}{c c c c c c c}
\hline
&$\beta_{imp}/\beta$ & $c_x$     & $c_t$    &$c$ &  $e/d$        &    $h$   \\
\hline
quench &6.4  & & &0.827(27)&0.954(1) & -0.12(3)\\
       & 6.45  & &  &0.814(89) &0.954(1) & -0.14(9)\\
\hline
MILC$_{2f}$&7.20 & 1.01(13)  & 0.74(4) &  0.94(13) & 0.9365(1) & -0.01(13) \\
MILC$_{3f}$&7.18 & 1.07(13)  & 0.76(3) &  0.99(16) & 0.9425(1) & -0.05(17) \\
\hline
\end{tabular}
\caption{The Kugo-Ojima parameter of the quenched $56^4$ lattice and that of the MILC$_{2f}$ and MILC$_{3f}$.  $c_x$ is the polarization along the spatial directions,  $c_t$ is that along the time direction,  $c$ is the weighted average of $c_x$ and $c_t$, $e/d$ is the trace divided by the dimension and $h$ is the horizon function deviation.}\label{kugo}
\end{table}


The running coupling in the $\widetilde{MOM}$ scheme in Landau gauge is given
by the product of the gluon dressing function and the ghost dressing function squared:
\[
\alpha_s(q)=q^6 D_G(q)^2 D_A(q).
\]
We fit the scale by comparing with the perturbative QCD(pQCD) result at the high momentum region. 

\DOUBLEFIGURE[h]{milcft_kgplt.eps,width=7cm}{alp_s_3fn.eps,width=7cm}{Kugo-Ojima parameter $u(0)$ of MILC$_{ft}$ configurations of $T/T_c=1.02$(blue diamonds),  $T/T_c=1.23$(red stars) and $T/T_c=1.32$(green triangles). } {The running coupling $\alpha_s(q)/\pi$ of MILC$_f$ in Landau gauge  The pQCD result of $N_f=3$ (upper dash-dotted line) and $N_f=2$ (lower dashed line) and the extraction of JLab are also plotted(blue boxes).}\label{alphaplt}

Recently an experimental extraction of an effective strong coupling constant was published from the Thomas Jefferson National Accelerator Facility(JLab) collaboration\cite{DBCK06}.   Using the Bjorken sum rule at 0 momentum, they predicted that the strong coupling constant approaches $\pi$. This prediction and data for $Q>0.4$GeV shown in Figure 2 are consistent with our lattice results. The infrared suppression in the Landau gauge does not agree with JLab result.

\section{The Lattice Coulomb gauge QCD}
We adopt the minimizing function 
$F_U[g]=||{\Vec A}^g||^2=\sum_{x,i}{\rm tr}
 \left({{A^g}_{x,i}}^{\dag}A^g_{x,i}\right)$, 
and solve $\partial_i ^gA_i({\Vec x},t)=0$ using the Newton method. We obtain $\displaystyle \epsilon=\frac{1}{-\partial D}\partial_i {A_i}$ from the eq. $\partial_i A_i+\partial _i D_i(A)\epsilon=0$.  Putting $g({\Vec x},t)=e^\epsilon$ in ${U^g}_i({\Vec x},t)=g({\Vec x},t)U_i({\Vec x},t)g^\dagger({\Vec x+i},t)$ we set the ending condition of the gauge fixing as the maximum of the divergence of the gauge field over $N_c^2-1$ color and the volume is less than $10^{-4}$,
$
Max_{x,a}(div A)^a(x) <10^{-4}
$.
This condition yields in most samples
\[
\frac{1}{8V}\sum_{a,x}({div A^a}_x)^2\sim 10^{-13}.
\]

We measure the color-Coulomb potential by first defining
\[
V_{Coul}^{ab}({x},{y})=\left\langle \langle a,x|{\mathcal M}^{-1}(-\partial^2){\mathcal M}^{-1}|b,y\rangle \right\rangle,
\]
where the outermost bracket denotes ensemple average, $a,b$ are color indices, $x=({\Vec x},t)$ and $y=({\Vec y},t)$.
We perform the Fourier transform
\begin{eqnarray}
&&V_{Coul}^{ab}({\Vec q})=\frac{1}{V}\sum_{{\Vec x},{\Vec y}}e^{-i{\Vec q}\cdot{\Vec x}}V_{Coul}^{ab}(x,y,t)e^{i{\Vec q}\cdot{\Vec y}}
=\left\langle \langle a,{\Vec q}|{\mathcal M}^{-1}(-\partial_\mu)\partial_\mu
{\mathcal M}^{-1}|b,{\Vec q}\rangle \right\rangle\nonumber\\
&&=\left\langle\sum_\mu \langle{\phi_\mu}^{a {\Vec q}}|{\phi_\mu}^{b {\Vec q}}\rangle\right\rangle
=\left\langle \sum_{\mu,x}{\phi_\mu}^{a,{\Vec q}*}(x){\phi_\mu}^{b,{\Vec q}}(x)\right\rangle\label{vcoul}
\end{eqnarray}
where we define
\begin{equation}
\frac{1}{\sqrt V}\sum_{y}|b,y\rangle e^{i{\Vec q}\cdot \Vec y}=|b, {\Vec q}\rangle
\end{equation}
and 
\begin{equation}
{\phi_\mu}^{b,{\Vec q}}(y)=\partial_\mu {\mathcal M}^{-1}|b,{\Vec q}\rangle
\end{equation}

We evaluate $\psi^b(y)$ as a solution of the differential equation for a plane wave source $\rho(y)$
\begin{equation}
(-\partial D)\psi^b(y)=\rho^b(y)
\end{equation}
and define ${\phi_\mu}^{b}(y)=\partial_\mu \psi^{b}(y)$. Following the DSE\cite{FeRe04}, we express the color-Coulomb potential as
\begin{equation}
V_{Coul}({\Vec q})=D_G({\Vec q}^2){\Vec q}^2f({\Vec q})D_G({\Vec q})
\end{equation}
where $D_G({\Vec q})$ is the ghost propagator and $f({\Vec q})$ is the Coulomb form factor.

\DOUBLEFIGURE[h]{f_dwf_2f3f.eps,width=7cm} {f_milcft123.eps,width=7cm}{The color-Coulomb form factor $f(q)$ of DWF $m=0.01$(green triangles), that of MILC$_{2f}$(violet stars) and MILC$_{3f}$(magenta diamonds).}{Same as Figure 3 but that of MILC$_{ft}$, $T/T_c=1.02$(blue diamonds), $T/T_c=1.23$(red stars)C
$T/T_c=1.32$(green triangles).}

The color-Coulomb form factors of MILC$_{2f}$, MILC$_{3f}$ and DWF $m=0.01$ shown in Figure 3 indicate that $\alpha_{Coul}({\Vec q}) $depends on the kind of fermion. In the case of DWF, we excluded exceptional samples whose $V_{Coul}({\Vec q})$ at zero momentum is extremely large ( there are a few such samples in DWF m=0.01 among 50 samples).
The results of MILC$_{ft}$ shown in Figure 4 indicate its temperature dependence.

\DOUBLEFIGURE[h]{DAts_2f3f.eps,width=7cm} {DAts_dwf0123.eps,width=7cm}{The Coulomb gauge transverse gluon propagator $D_A(q)$ of MILC$_{2f}$ (violet stars) and MILC$_{3f}$(magenta diamonds).}{The Coulomb gauge transverse gluon propagator $D_A(q)$ of DWF$m=0.01$(green triangles), $m=0.02$(magenta diamonds) and $m=0.03$(orange stars).}

The gluon propagator obtained as an average of data measured at each time slice  does not depend much on the kind of fermion and temperature as shown in Figure 5 (MILC$_{2f}$ and MILC$_{3f}$) and in Figure 6(DWF).

In the pQCD region
the running coupling is defined from the color-Coulomb potential $\alpha_{Coul}({\Vec q})=\frac{11N_c-2N_f}{12N_c}q^2V_{Coul}({\Vec q})$. In the infrared region, we measure the running coupling in Coulomb gauge as, 
\[
\alpha_I({\Vec q})={\Vec q}^5 D_G({\Vec q})^2 D_A({\Vec q})
\]
where $D_A({\Vec q})$ is the
gluon propagator.

The gauge field $A_{x,\mu}$ and the link variable $U_\mu(x)$ are related by,
$U_\mu(x)=e^{\lambda^a A^a_{x,\mu}}$ in \cite{Lu98}, but $U_\mu(x)=e^{\Lambda^a A^a_{x,\mu}}$ in our case, where $\displaystyle \Lambda^a=\frac{\lambda^a}{\sqrt 2}$ with Gell-Mann's SU(3) generator $\lambda$.

Conformance with the unit length transporter $e^{\frac{\lambda^a}{2}A^a_\mu(x)}$ in the continuum theory yields the gluon dressing function $D_A(q)$ that approaches 2 in large ${q}$.  
Conformance with the transporter $e^{\Lambda^a A^a_\mu(x)}$ i.e.
the normalization of $A_{x,\mu}$ such that $D_A({\Vec q})$ approaches 1 yields the $\alpha_I({\Vec q})$ consistent with the JLab result. We multiply $1/2$ to $D_A({\Vec q)}$ in the calculation of $\alpha_I({\Vec q})$.

\DOUBLEFIGURE[h] {alpIts_2f3f.eps,width=7cm}{milcft_alpItsl135.eps,width=7cm}{The running coupling $\alpha_I(q)/\pi$ of MILC$_{3f}$ and MILC$_{2f}$ in Coulomb gauge.  The pQCD result of $N_f=3$ (upper dash-dotted line) and $N_f=2$ (lower dashed line) are also plotted.}{The running coupling $\alpha_I({\Vec q}^2)/\pi$ of MILC$_{ft}$. $T/T_c=1.02$(blue diamonds), $T/T_c=1.23$(red stars) and $T/T_c=1.32$(green triangles).}\label{alphaplt2}

\smallskip
\FIGURE[br]{\epsfig{file=alpIts_dwf0123l.eps,width=7cm} \caption{The running coupling $\alpha_I({\Vec q}^2)/\pi$ of DWF $m=0.01$(green triangles),0.02(magenta diamonds) and 0.03(orange stars).}\label{alp_dwf}}

The running coupling $\alpha_I({\Vec q})$ of  MILC$_{2f}$ and MILC$_{3f}$ are shown in Figure 7 and that of MILC$_{ft}$ is shown in Figure 8.

We observe that $\alpha_I({\Vec q})$ of MILC$_{2f}$, MILC$_{3f}$ and DWF (Figure 9) agree with the extraction of JLab.  We emphasize that 
the adjusted parameter is only the normalization of the gauge field which
appear in the transporter as a coefficient of $\Lambda$. 

In case of zero temperature quenched configuration, the Kugo-Ojima color confinement parameter $c$ is about 0.8. In case of unquenched configurations it is consistent with 1 at zero temperature and decreases as temperature rises.

The $A^2$ condensate is finite and ghost condesate parameters is small at zero temperature, but near $T_c$ they are consistent with 0.
The gluon propagator in the Landau gauge and in the Coulomb gauge are infrared finite. Its large volume limit is yet to be investigated.

The running coupling of the Coulomb gauge $\alpha_I({\Vec q})$ may approach $\pi$ at zero momentum, independent of the kind of fermion. 

The infrared suppression of the running coupling in Landau gauge may be due to the singularity of the color anti-symmetric ghost propagator that disturbs the color diagonal ghost propagator.
Quark has the effect of magnifying the square norm of the color anti-symmetric ghost propagator and reduces its fluctuation\cite{FuNa06c}.
 We observed that the color diagonal ghost propagator is temperature independent. Color anti-symmetric ghost propagator of a system containing dynamical quarks is temperature dependent\cite{FuNa06d}.

Presence of dynamical quarks is crucial in defining infrared features of the QCD. Lattice data always suffer from finite size effect. Comparison with theoretical approaches as Dyson-Schwinger equation\cite{FiAl03} is important. Calculation of the effective coupling in other scheme and comparison with experimental data are also challenging issues\cite{BrLu95, BMMR02}.

\vskip 0.3 true cm
The numerical simulation was performed at KEK using Hitachi-SR11000, at Yukawa institute of theoretical physics of Kyoto University using NEC-SX8 and at CMC of Osaka University using NEC-SX8.
This work was supported by the KEK supercomputing project No.5(FY2006).
We thank Alexandre Deur for sending the results of JLab,
the MILC collaboration for supplying very useful data, and the BNL lattice QCD group for the information on the gauge
configurations of the QCDOC collaboration.

\end{document}